\documentclass[oldversion]{aa}
\usepackage{txfonts}
\usepackage[dvips]{graphicx}
\usepackage{natbib}
\bibpunct{(}{)}{;}{a}{}{,}
\defcitealias{abl05}{Paper I}

\begin{document}
\authorrunning{M. Axelsson et al.}
\title{Probing the temporal variability of Cygnus X-1 into the soft state
\thanks{Table 1 is only available in electronic form at the CDS via
anonymous ftp to cdsarc.u-trasbg.fr (130.79.125.5) or via
http://cdsweb.u-strasbg.fr/Abstract.html}}
\author{Magnus Axelsson \and Luis Borgonovo \and Stefan Larsson}
\institute{Stockholm Observatory, AlbaNova, SE- 106 91 Stockholm, Sweden}
\offprints{M. Axelsson,\\ \email{magnusa@astro.su.se (MA), luis@astro.su.se (LB)}}
\date{Received 21 October 2005 / Accepted 9 March 2006}

\abstract{Building on results from previous studies of Cygnus~X-1, we
analyze \textit{Rossi X-ray Timing Explorer} (RXTE) data taken when
the source was in the  soft and transitional spectral states. We look 
at the power
spectrum in the 0.01 -- 50 Hz range, using a model consisting 
of a cut-off power-law and two Lorentzian components. We are able to
constrain the relation between the characteristic frequencies of the
Lorentzian components, and show that it is consistent with a power-law
relation having the same index (1.2) as previously reported for the
hard state, but shifted by a factor $\sim2$. Furthermore, it is shown
that the change in the frequency relation seen during the transitions can be
explained by invoking a shift of one Lorentzian component to a higher
harmonic, and we explore the possible support for this interpretation
in the other component parameters. With the improved soft state
results we study the evolution of the fractional variance for each
temporal component. This approach indicates that the two Lorentzian
components are connected to each other, and unrelated to the
power-law component in the power spectrum, pointing to at least two
separate emission components. }

\maketitle

\section{Introduction}

\object{Cygnus X-1} is one of the most studied X-ray sources, and is
often quoted as the prototype black hole binary system. The source
exhibits two main spectral states, commonly referred to as hard and
soft, with a brief intermediate state during transitions
\citep[e.g., ][]{zdz02,zg04}. Recently, a study of the broad-band
spectra of Cyg~X-1, mainly in the hard state, has been presented by
\citet{ibr05}, and a comprehensive study of all states of the source
in the \mbox{3--200 keV} range has been carried out by \citet{wil05}. Several 
models have been proposed to explain the observed states and transitions.
The two main components of such models are usually a geometrically
thin, optically thick accretion disk and a hot inner flow or
corona. The models vary in the geometry and properties of mainly the
corona/Comptonizing region \citep[for a review on coronal models,
especially in regard to timing characteristics,
see][]{pou01}. However, changes in the inner radius of the accretion
disk are a common source of variability in many models
\citep[e.g.,][]{pou97,esi98,chu01,zdz02}.

The changes in spectral state are evident also in the power density
spectrum (PDS). In the hard state the PDS is characterized by a
flat-topped component up to $\sim 0.2$ Hz, where it breaks to a
$f^{-1}$ slope that steepens at a few Hz
\citep[e.g.,][]{bh90,now99}. In the soft state the PDS is dominated by
a $f^{-1}$ component that steepens around 10 Hz
\citep[e.g.,][]{cui97a}. The 1996 state transition also revealed an
intermediate PDS, showing a $\sim f^{-1}$ slope at lower frequencies,
a flatter component around $0.3-3$ Hz, and above that similar to the
hard state PDS. In many observations there is evidence for a
quasi-periodic oscillation (QPO) at a few Hz \citep{bel96,cui97b,cui99}.

Previous studies of the PDS of Cyg~X-1 have shown correlations between
different temporal features as well as between temporal and spectral
components \citep[e.g.,][]{gil99}. Correlations between features in the
hard state PDS were first reported in \citet{wvdk99}. \citet{now00} 
showed that the hard state PDS of Cyg~X-1 could be well fit using 
Lorentzian components. This was expanded in a large scale study of the
hard state PDS conducted by \citet{pot03}, who also included some `failed 
state transition' PDS. A systematic study of the PDS for
all spectral states of Cyg~X-1 was conducted by \citet[][ hereafter
Paper I] {abl05} using archival data from the \textit{Rossi X-Ray
Timing Explorer} (RXTE) satellite. Using a model consisting of a 
cut-off power-law and two Lorentzian profiles, we were able 
to fit the
majority of PDS, and follow the components from hard state through the
transitions and back. While the results showed that the Lorentzian
components were present in a significant fraction of the soft state
PDS, their behavior was not well constrained. In this paper we
reanalyze the soft state data and expand it with observations from an
additional soft state in May-August of 2003, recently made
available. With the improved soft state results we are able to
decompose the PDS and study the contribution from each component
separately.

We begin this paper with a brief description of the data used, the
analysis procedure and the model components (Sect.~\ref{analysis}).
This is followed by a presentation of our results in Sect.~\ref{results}.
In Sect.~\ref{discuss} we discuss our findings, and put them in context
of other results. Finally our findings are summarized in Sect.~\ref{conc}.
  
\section{Observations and Data Analysis}
\label{analysis}
The data and methods of analysis used in this paper are in essence
the same as in \citetalias{abl05}. Therefore we only give a brief 
summary of the data and extraction process, with emphasis on the
new data and the parts of the analysis that have been improved. For
more details we refer to \citetalias{abl05}.  

\subsection{The archival data}
Since its launch on December 30, 1995, the RXTE satellite has observed 
Cyg~X-1 in more than 750 pointed observations, with a typical on-source 
time of 3 -- 4 ks. During this time, the
All-Sky Monitor (ASM) instrument on board the satellite has also
provided a continual coverage of the source in the 2--12 keV range. 

As in our previous study, the data are obtained
with the Proportional Counter Array \citep[PCA,][]{jah96} on RXTE,
which consists of five identical Proportional Counter Units (PCUs),
and is sensitive in the 2--60 keV energy range. The new soft state
roughly spans the time from May 2003 to August 2003. We
include all pointed observations made of the source during these four
months, $\sim 4$ days per month. Note that some of these data
are presented in \citet{pot05}. The data mode used was the `Generic
Binned' mode. We also calculate count (hardness) ratios, using data from
the Standard2 configuration.
 
Lightcurves were extracted with the standard RXTE data analysis 
software FTOOLS, version 5.2, using standard screening criteria: a
source elevation $>10^{\circ}$, a pointing offset $<0\fdg01$ and a
South Atlantic Anomaly exclusion time of 30 minutes. The energy
range of the data used for the PDS is $\sim 2-9$ keV, with the soft
and hard bands of the count ratio taken as $2-4$ keV and $9-20$ keV
respectively. For the PDS, the lightcurves were extracted with a time
resolution of 10 ms. The lightcurves used in determining count ratios
were extracted with 16 s time resolution. All lightcurves were
normalized to one average PCU using the FTOOL \texttt{correctlc}.

\subsection{Calculating the power spectra}

The modelling of the soft state PDS in \citetalias{abl05} (e.g., Fig.~9) 
showed the need to extend the studied frequencies above 25~Hz. This requires
proper treatment of dead-time effects, and we therefore correct the
estimated Poisson level. We use the correction for general paralyzable
dead-time of \citet{zha95} as presented in the first corrective
term in Eq.~6 of \citet{jer00}. We chose an upper limit of 50 Hz.

To calculate the power density spectra the high-resolution light
curves were divided into segments of $2^{13}$ bins ($\sim 82$ s). The
resulting PDS were then gathered into groups of 10 and each group
averaged into one final PDS. Each PDS thus covers $\sim 15$ minutes of
observation time, allowing changes occuring on this timescale to be
studied. As shown in \citetalias{abl05}, significant changes in the PDS 
of Cyg~X-1 do occur on these short timescales. The PDS have been rebinned
using semi-logarithmically spaced bins.

The normalization used for the PDS is that of \citet{bh90} and 
\citet{miy92}, where the integral over a frequency range gives the 
square of the fractional root-mean-square (RMS) of that frequency
interval. In addition, the PDS are plotted in units of frequency times
power versus frequency ($fP_f$ versus $f$).  Together with the new
data from 2003, the soft state data from \citetalias{abl05} were
reanalyzed and corrected for dead-time effects, improving the
signal-to-noise ratio. Altogether, a total of 614 soft state PDS 
were obtained. 
 
\subsection{Modeling}
\label{models}

\citet{now00} showed that the PDS of Cyg~X-1 in the hard state 
could be successfully fit using Lorentzian components. The model was extended
by \citet{pot03}, who added a power-law component in order to fit the
lower frequencies, especially during the `failed state transitions'. 
Building on these results, it was shown in \citetalias{abl05}
that by introducing a cut-off power-law component together with two
Lorentzians, we were able to provide a good fit to the PDS of Cyg~X-1 in 
all states of the source. We could thereby for the first time model the
complete evolution of the PDS. As the parameters of that model differ
somewhat from those in the previous studies, we present their
definition again here. The Lorentzian profiles have the form
\begin{equation}
L_i(f)=\frac{H_iW_i\nu_i}{(f-\nu_i)^2+fW_i\nu_i} 
\label{loreq}
\end{equation}
\noindent where $H_i$ is the value of $fP_f$ at the peak frequency $\nu_i$. 
The dimensionless parameter $W_i$ gives a measure of the relative 
width of the
profile in the $fP_f$ representation. The parameterization is chosen
such that for a given $W_i$ and $H_i$, the integral of the Lorentzian
(and thus its fractional contribution to the total RMS variability) is
the same for any value of $\nu_i$. The cut-off power-law necessary in
the transitional and soft states is of the form
\begin{equation}
Pl(f)=Af^{-\alpha} e^{-f/f_c}
\label{pleq}
\end{equation}
Where A is the normalization constant, $\alpha$ is the slope, and
$f_c$ is the turnover frequency. Of the 614 soft state PDS obtained
from the observations, 143 allowed a fit using all three components of
the model, one or both Lorentzian components being too weak to
constrain in the remaining PDS. These fits were first performed with
all parameters free, and then with the width of the first Lorentzian,
$W_1$, frozen to a value of 0.6. While freezing $W_1$ did not
affect the number of PDS where $L_2$ was detected, it allowed us to
better constrain $L_2$ in those cases and the results of these latter
fits were therefore kept. Table 1 lists the results of all fits
used in this paper, including the ones from \citetalias{abl05}, and is
available at the CDS in electronic form only. The first six columns
give the proposal number and sub-ID (e.g., ``60090-01-39-00''), the start 
and stop times in MJD of the lightcurve used to calculate the PDS, a
number identifying the model components present in the fit, and
the average 2--9 keV flux and (9--20 kev)/(2--4 keV) hardness during 
that time. This is followed by 18 columns giving the parameter value 
and error for the three parameters of the power-law ($A,\; \alpha,\; f_c$) 
and the parameters of each of the two Lorentzians $L_1$ and $L_2$ 
($\nu_i$, $H_i$, and $W_i$).

The model used is not aimed at giving the fullest possible 
description of the PDS of Cyg~X-1. For example, we do not attempt to 
model very small features of the PDS, and indeed the signal-to-noise
ratio is insufficient for us to accurately do so. The strength of the
model lies rather in its simplicity, allowing us to accurately fit the 
main characteristics of the PDS in all states, and track changes 
occuring on short timescales. Including a cut-off to the power-law
proved essential in allowing the same model to fit {\it all} states, 
thereby giving a more complete picture of the power spectral evolution.

\subsection{Definitions of state}
\label{states}

When determining the state of Cyg~X-1, it is possible to 
look at characteristics of either radiation spectra or temporal analysis. 
Current definitions of state, derived from either of these characteristics,
are mainly phenomenological. These definitions all agree on the `classical'
hard and soft states. However, difficulties arise when attempting to
define the extent of these states, and the boundaries of the transitional
(or intermediate) state. Until the physical processes behind the state
transitions are understood, these definitions will retain some of their
arbitrary nature.

In an extensive study of the radiation 
spectra of Cyg~X-1 in all states, \citet{wil05} showed that there is 
a continuous spectral evolution between the states of the source. 
While timing spectra are generally able to provide sharper criteria,
it was shown in \citetalias{abl05} that there is a continuum of
PDS between the hard and soft states, with examples covering all stages 
of the evolution. Therefore, criteria of state such as the temporal 
features crossing certain frequency boundaries, or the appearance of a
power-law component, do not by themselves provide good indicators of
state. However, our previous study also revealed that several
parameter correlations change behavior at the same stage of the evolution, 
and we therefore regard these changes to be a natural marker of the source 
entering the transitional state. Studies
have also shown decreased coherence and enhanced time lags between the
hard and soft lightcurves during transitions, both in Cyg~X-1
\citep{cui97b,pot00} and other sources \citep[such as \object{GX~339-4} 
and \object{XTE~J1650-500}, see ][ respectively]{now02,kal03}. \citet{ben04}
use a combination of photon index $\Gamma$ and time lags between the 
$2-4$ keV and $8-13$ keV channels to define the state of Cyg~X-1. 
In their long term study of the hard state, \citet{pot03} showed that the 
state is not uniform, and a distinction is made between `quiet' and `flaring'
hard states, where flares and `failed state transitions' are more
frequent in the latter. The classification of state is thus not rigorously 
defined, and remains
an area of active research \citep[for a recent review, see][]{zg04}.

For consistency with \citetalias{abl05}, we have used the same criteria
when defining state, based on the changes seen in the temporal components.
With this definition, the PDS in the hard state only show $L_1$ and $L_2$.
As the source enters the transition, the power-law component enters the
frequency window, and several parameter relations change behavior.
In the soft state, $L_2$ is not seen or significantly weaker than $L_1$,
or the PDS is completely dominated by the cut-off power-law component. 
For comparative purposes, we note that our definition is consistent
with a criterion of the effective spectral index ($3-12$ keV) 
$\Gamma \ga 2.4$ for the soft state used elsewhere 
\citep[e.g., ][]{zdz02}. 

During the `quiet' hard state, there are 
periods when a third component appears at the higher frequencies of the 
PDS and the hardness increases. We refer to these episodes as the
`hard edge of the hard state', and while we do not model them in this 
paper, they are part of the data from \citetalias{abl05} presented in
Sect.~\ref{evolution}.  

\section{Results}
\label{results}

We now turn to the results of our analysis, starting with the study of
the soft state data. Following this analysis, we show that a
decomposition of the PDS is now possible, and present the evolution of
each temporal component.

\subsection{The two Lorentzians}
\label{compres}

\begin{figure}
\resizebox{\hsize}{!}{\includegraphics{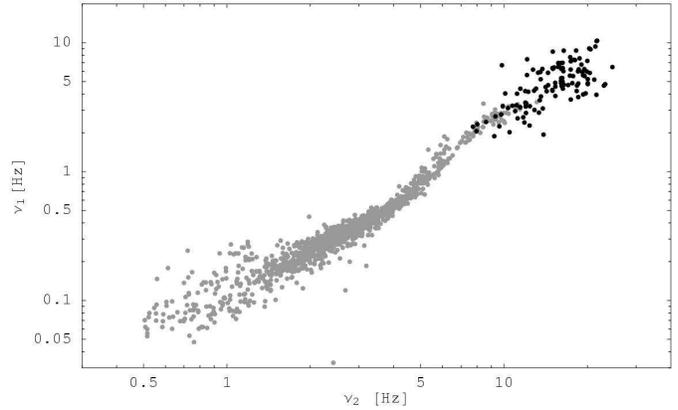}}
\caption{Relation between the two peak frequencies of the Lorentzian 
components. The grey points are 1414 hard state and transition points 
from \citetalias{abl05}, while the black points are 143 soft state 
points resulting from the analysis in this paper.}
\label{freqrel}
\end{figure}

\begin{figure*}
\centering
\includegraphics[width=14cm]{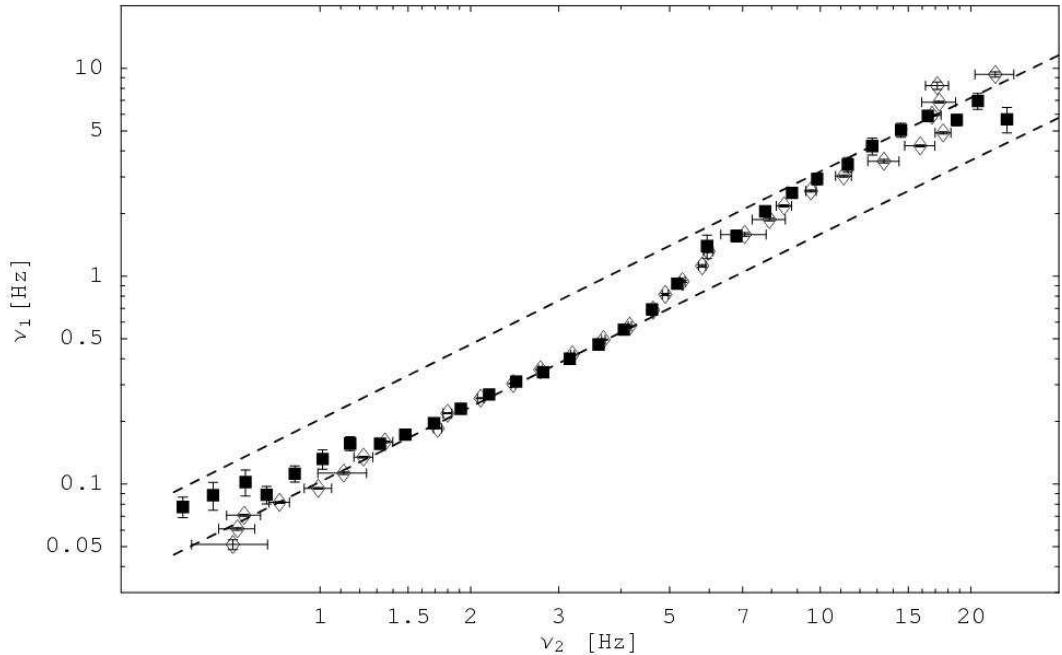}
\caption{Results of binning in $\nu_1$ (open diamonds) and $\nu_2$ (solid 
squares). The lower line is the same as in Fig.~17 of \citetalias{abl05},
a fit to the hard state points (index $1.20\pm0.01$), and the upper one 
a power-law with the same index ($1.19\pm0.14$) fit to the soft state 
points (binning along $\nu_2$). Note how the binning along $\nu_1$ 
deteriorates as the source enters the soft state, but follows the power-law 
in the hard state to the lowest frequencies. The binning along $\nu_2$ 
deviates from the hard state fit at the lower frequencies, but extends 
the relation in the soft state to higher frequencies.}
\label{binrel}
\end{figure*} 

While the peak frequencies of the two Lorentzian components follow a 
power-law correlation in the hard state, the relation in the soft state
is still uncertain \citepalias[][ Fig.~9]{abl05}. As this deeper
study is partly aimed at resolving this behavior,
we begin by showing the frequency relation in Fig.~\ref{freqrel}.
As can be seen in the figure, the spread of points is quite large in
the soft state (black points). To determine whether there is a
significant trend also in the soft state we will attempt to bin the 
data points. The problem with the binning is the same as in the more 
common case of regression, when choosing which parameter to treat as 
the independent one. In order to determine the effect of the binning
procedure in this particular case we first performed Monte Carlo
simulations, where we introduced noise to points following a power-law
relation with known index. The resulting distributions were then
binned along both axes and the indices determined. As long as the
uncertainties are small, binning the data recovers the correct
index. However, the results clearly show that an increase in noise
tends to produce an artificial flattening effect along the axis of
binning. If the amount of noise varies between the parameters, the
best results are achieved when binning is done along the parameter
with smallest errors.  Returning to the analogy with regression, this
amounts to the parameter with the smallest errors being chosen as the
independent one. In addition, as our frequency range is limited, there
are border effects as the components approach these limits, causing
a biased distribution in the frequency nearest the border.

We now apply the same binning procedure to the data points of
Fig.~\ref{freqrel}. In the soft state, the largest spread is along
$\nu_1$. Therefore, binning along $\nu_2$ more accurately reveals the
behavior of the relation. Conversely, for the hard state data, it
is the spread along $\nu_2$ that dominates. In this range, binning along
$\nu_1$ gives a better indication of the behavior of the
relation. Figure \ref{binrel} shows the result for both binnings.   
Also included in the figure is the power-law fit to the hard
state points made in \citetalias{abl05}, with index $1.20 \pm 0.01$,
as well as a power-law fit instead to the points in the soft state 
(solid squares). Between the extremes, the results of the two
binnings are indistinguishable. The flattening effects seen in
Fig.~\ref{binrel} as the components approach the frequency limits are in 
agreement with the results of our simulations. Our change of binning 
parameter is thus well founded, but the outermost points at either extreme 
should be treated with caution.

With the better soft state data and the binning along $\nu_2$, it is
clear from Fig.~\ref{binrel} that the relation in the soft state
follows a power-law with an index similar to that of the hard
state. The result is of course dependent on the where the soft state
region is assumed to begin. Setting the lower limit too high in
frequency ($\nu_2 \ga 10$~Hz) results in a large uncertainty as the
dispersion is high in this region. Lowering the boundary to include
the asymptotic behavior of the transition ($\nu_2 \ga 7$~Hz) reduces
the fitting error, but introduces a systematic effect increasing the
index. For all reasonable choices of this boundary (i.e., \mbox{$7$~Hz
$\la \nu_2 \la 10$~Hz}) an index of 1.2 is consistent with the result
of the fit. We will assume a conservative estimate of \mbox{$1.19 \pm
0.14$} as the index for the soft state points, with the value
derived from the fit with boundary at 8.8 Hz, and the error reflecting the
variations arising from the range of possible boundary choices. The
form of the power-law used for these fittings is:
\begin{equation}
\nu_1=B\;\nu_2^\beta
\end{equation}
\noindent where $B$ is a constant and $\beta$ the power-law index.

Adopting the fits for the two states, a look at the constants 
of the power-laws 
reveals the values of $B=0.103 \pm 0.002$ in the hard state and 
$B=0.19 \pm 0.05$ in the soft state. The ratio of the constants is close
to 2, and together with the fitted indices a shift of the hard state relation 
by a factor of two is consistent with the data. Is it therefore possible 
that the break in the relation during the transitions is simply a shift 
to a higher harmonic? In order to accurately model such a shift, three 
components would be necessary during the transitions. As our model only
contains two components, it would not catch the change if the shift
occurs gradually. Unfortunately, attempting to fit the PDS with three
Lorentzian components shows that a third component is impossible to
constrain in the data, and rarely produces a better fit. We therefore
attempt a different approach using simulations.    

\subsection{Simulations}
\label{sims}

In order to test the hypothesis of $L_1$ shifting to its first harmonic,
we conducted a series of Monte Carlo simulations where such an effect
was included. Artificial power spectra were created and fitted with our
model of two Lorentzian components plus a cut-off power-law. As input
parameter for the simulations we use $\nu_2$, the peak frequency of
$L_2$. This is to get the same distribution of frequency points as in
the observed data. For the other parameters we utilize the parameter
correlations found in
\citetalias[Figs.~10-13 of~][]{abl05}. We use analytic expressions 
to approximate these behaviors (shown in Fig.~\ref{parrels}), adding
some random Gaussian spread. For each $\nu_2$ a corresponding $\nu_1$
was derived assuming the hard state relation between the frequencies
(lower line in Fig.~\ref{binrel}), and values for $H_2$ and $W_2$
were calculated. Finally, a third Lorentzian component, $L_1^+$, was added
with $\nu_1^+=2 \nu_1$, i.e. the first harmonic of $L_1$. The
parameters $W_1$, $W_1^+$, $H_1$ and $H_1^+$ were now determined from
the relations seen in the left panels of Fig.~\ref{parrels}. As $W_1$ is 
measured to center around one value at lower frequencies, and a lower value at
higher frequencies, these different values were adopted for $W_1$ and
$W_1^+$ (solid and dashed lines in panel \textbf{b} of
Fig.~\ref{parrels} respectively). To simulate a shift in power from
$L_1$ to $L_1^+$ during the transitions, the analytical approximation
shown in panel \textbf{a} of Fig.~\ref{parrels} was multiplied by
modulating factors:

\begin{figure}
\resizebox{\hsize}{!}{\includegraphics{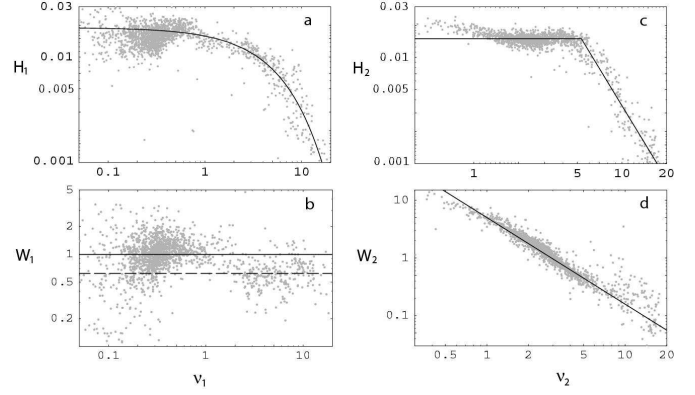}}
\caption{The parameter correlations from \citetalias{abl05} (gray points)
and the analytical expressions used to approximate them in the simulations
(black lines). The two left panels show $H_1$ (panel \textbf{a}) and $W_1$ 
(panel \textbf{b}) as a function of $\nu_1$. The two right-hand panels
show the correlations for $H_2$ and $W_2$ versus $\nu_2$ (panels \textbf{c}
and \textbf{d}). In panel \textbf{b}, the black line indicates the 
approximation used for $L_1$ and the dashed line for $L_1^+$. Note also
the deviation from the approximation at the lowest frequencies in panel 
\textbf{d}. These points are from the hard state, when $L_1$ approaches
the lower frequency boundary, and when the hardness is highest. As our
simulations mainly cover the transition and soft state, this deviation
does not influence our results. }
\label{parrels}
\end{figure}

\begin{figure}
\resizebox{\hsize}{!}{\includegraphics{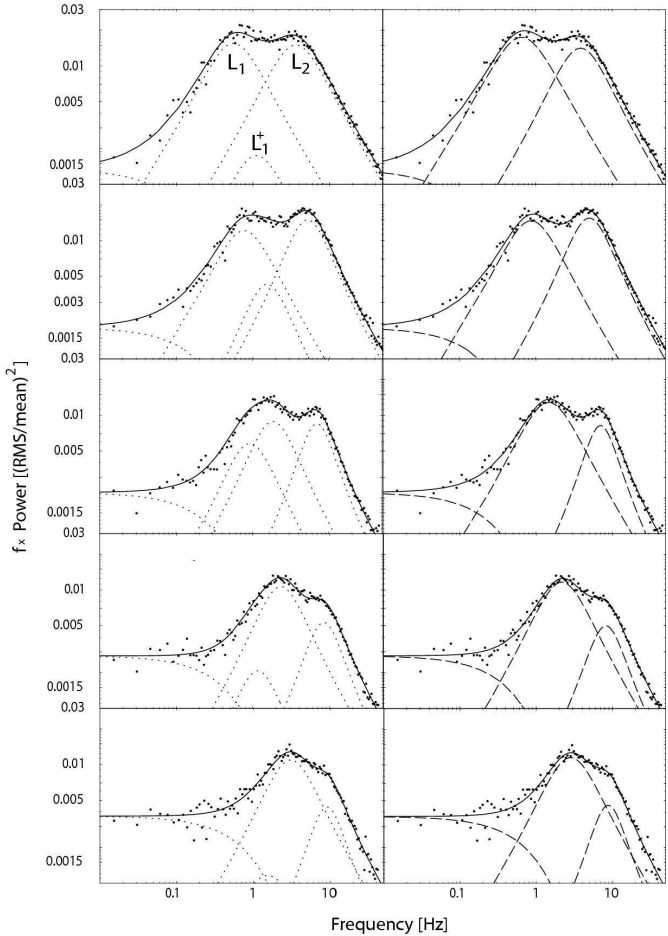}}
\caption{A sequence of five simulated power density spectra, displaying 
the evolution from hard state to the soft state (top to
bottom). The panels on the left show the simulated model components
(dotted lines): a lower frequency Lorentzian $L_1$ and its first
harmonic $L_1^+$, a high frequency Lorentzian $L_2$ and a cut-off
power-law. The solid line is the sum of the noise free components, the
points indicate the result after the addition of noise. The panels on
the right show the results of fitting the simulated PDS with our
model, using only two Lorentzians and a cut-off power-law (dashed
lines). The solid line is the sum of the components. Note that even
when $L_1$ and $L_1^+$ are of nearly equal strength (e.g., third row
from top), they are not resolved in the noise free PDS (solid line in
left panel), and the sum is well fit by one wider Lorentzian
component.}
\label{simex}
\end{figure}

\begin{equation}
m_1(\nu_2) =\frac{1}{1 + \exp[-\frac{\nu_t - \nu_2}{\Delta \nu}]}\\
m_1^+(\nu_2) =\frac{1}{1 + \exp[+\frac{\nu_t - \nu_2}{\Delta \nu}]},
\label{modeq}
\end{equation}

\noindent where $m_1(\nu_2)$ is the factor for $H_1$, $m_1^+(\nu_2)$ the 
factor for $H_1^+$, $\nu_t$ the value of $\nu_2$ roughly in the middle
of the transition \mbox{(6~Hz $\la \nu_t \la 7$~Hz)} and $\Delta \nu$
is a constant related to the width of the transition ($\sim 1$~Hz). The 
values of these parameters are constrained by the
transition region as seen in Fig.~\ref{binrel}. The figure shows that
the influence of $L_1^+$ must start at $\nu_2 \sim 4$~Hz and
completely dominates above $\nu_2 \sim 8$~Hz. In our simulations,
$\nu_t$ was set to 6.5~Hz and $\Delta \nu$ to 1~Hz.

\begin{figure}
\resizebox{\hsize}{!}{\includegraphics{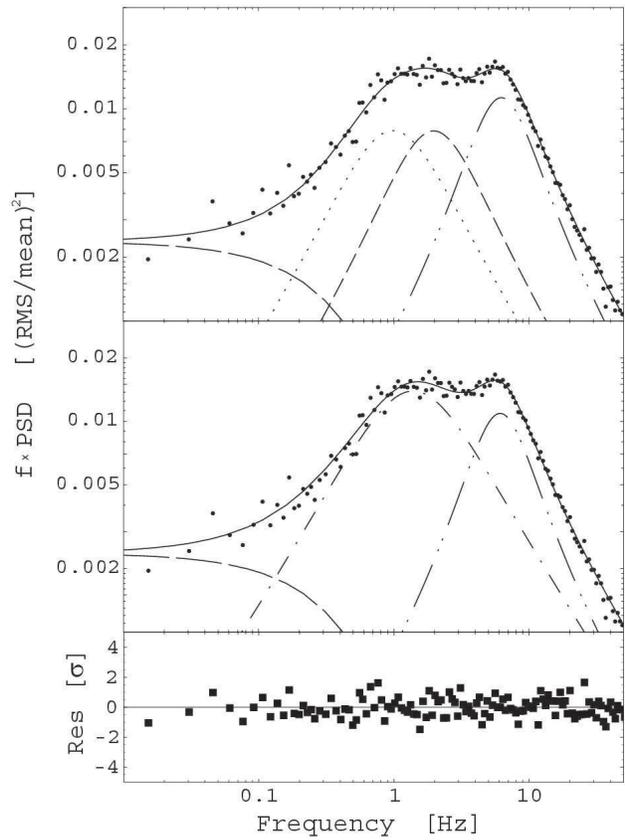}}
\caption{Result of fitting an artificially created PDS with the same model used
for the data. The simulated PDS is shown in the top panel, with the power-law
component as the long-dashed line, $L_1$ dotted, $L_1^+$ dashed and $L_2$
dot-dot-dashed. The middle panel shows the result of the fit using only
two Lorentzian components, $L_1$ (dot-dash) and $L_2$ (dot-dot-dash). The
bottom panel shows the residuals of the fit. Note that there is no trace
of the third component. Trying to add a third Lorentzian to the
fit is not feasible, as the data do not allow it to be constrained.}
\label{simfit}
\end{figure}

Finally, a power-law component was added and random Gaussian noise was 
added to the resulting PDS to account for the uncertainty of the
measured PSD values, and thereby giving the same signal-to-noise ratio as in
the observational data. We use Gaussian noise as the PDS are the results 
of both averaging and rebinning, making the fluctuations approximately 
Gaussian. This assumption is least valid at lower frequencies, as each bin 
here is the average of only a few points. However, these frequencies are 
not of interest in this context. Examples of resulting PDS are shown in the
left column of Fig.~\ref{simex}, along with the result of the fitting
(right column). The difficulty of resolving the two components is very
clear in the figure. As the width of each Lorentzian is very large
compared to their separation, they are unresolved even in the pure
signal (solid black line in the left-hand panels). Thus, even with a 
large increase in signal-to-noise ratio it will not be possible to
determine directly whether these two components are in fact present 
in the data. This becomes more clear in Fig.~\ref{simfit}, which shows
the result of the fitting routine when $L_1$ and $L_1^+$ are equally
strong, reflecting the most favorable scenario. From the residuals of 
the fit (bottom panel), it is clear that a single Lorentzian can 
successfully replace both $L_1$ and $L_1^+$.

\begin{figure}
\resizebox{\hsize}{!}{\includegraphics{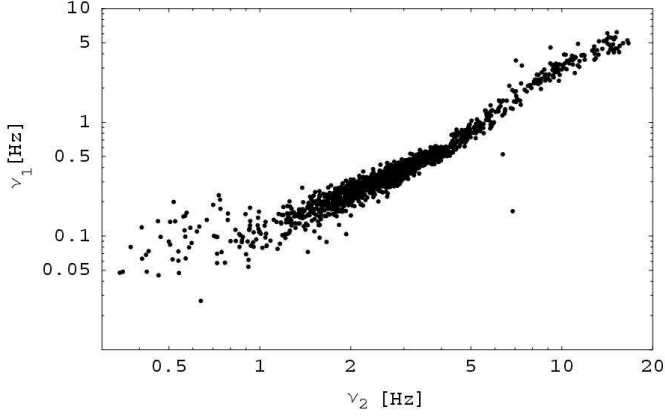}}
\caption{Results of fitting artificially created PDS with the same model used
for the data. In the artificial PDS a shift from $L_1$ to its first harmonic
is introduced, successfully reproducing the behavior in the data. Note that
the general trends of Fig.~\ref{freqrel} are also reproduced: increased
relative spread towards the lowest frequencies, low relative dispersion 
during the transition and increased relative spread in the soft state.}
\label{simfreqs}
\end{figure}

The artificial PDS were put through the same fitting routine
as the observed data, and the parameters of the resulting fits (right-hand 
panels in Fig.~\ref{simex}) plotted in the same way. Fig.~\ref{simfreqs} 
shows the resulting frequency correlation. Fitting the artificial PDS with
the model used for the data produces results nearly identical to those
obtained from the observational data. Not only is the change in
behavior around the transition ($\nu_2 \sim 4-9$~Hz) the same, but
when comparing with Fig.~\ref{freqrel} it is apparent that also the
general trend of the data points is reproduced. The relative
dispersion is greater during the soft and hard states, with a minimum
occuring during the transitions. As the components move to lower
frequencies, the points in Figs.~\ref{freqrel} and \ref{simfreqs} both
show the same gradual increase in relative dispersion. However, the
dispersion of the soft state points is smaller in the simulations than
in the data. This difference can be traced to the cut-off of the
power-law component. As the power-law extends through most of the
frequency range in the soft state, placing the cut-off too high in
frequency will capture power from $L_2$. Since $L_2$ is weak in the soft
state, this artificial weakening increases the uncertainty in
frequency. In our simulations, we can use our knowledge of the
power-law to eliminate this effect, thus reducing the dispersion in
the soft state. When the parameters of the power-law are left
completely free also in our simulations, the scatter of the soft state
points is significantly increased.

It should be noted that the results do not change significantly if the
amount of noise introduced is changed, nor does it depend to any large
extent on the exact expression used to approximate the parameter
relations.  Perhaps not surprisingly, the results are most sensitive
to the modulation function used to vary the strengths of $L_1$ and
$L_1^+$. This is evident in the $H_1-\nu_1$ relation (panel \textbf{a}
in Fig.~\ref{parrels}). If $L_1$ starts to decline while $L_1^+$ is
still very weak, the fitted $H_1$ values will show a dip at the
corresponding frequency, and there is no such feature present in the
data. We have tested expressions with both asymptotic exponential
behavior and behavior of power-laws of varying index.
The conclusion is that the rise of $L_1^+$ and the decay of $L_1$ must 
be fairly swift. This is also evident from the data in Fig.~\ref{freqrel}, 
as the transition occurs over a fairly narrow frequency range. We found
that the observational results were most readily reproduced using
exponential functions, as in Eq.~\ref{modeq}. Using, e.g., polynomial
modulation functions will achieve good agreement with the frequency
relation of Fig.~\ref{freqrel}, but cannot successfully reproduce the
other parameter relations (gray points in Fig.~\ref{parrels}). In 
particular, the inadequacy in correctly reproducing the parameter 
correlations was seen also for modulations of the functional form 
$\left(1+\left(\nu/\nu_0\right)^2\right)^{-1/2}$, typically associated
with band-pass filters.

The results from the simulation reproduce the behavior observed from the
data for all parameter correlations. The interpretation of a 
shift in power from $L_1$ to its first harmonic during hard to soft
transitions is therefore consistent with the data. 

\subsection{Evolution of the PDS}
\label{evolution}

\begin{figure}
\resizebox{\hsize}{!}{\includegraphics{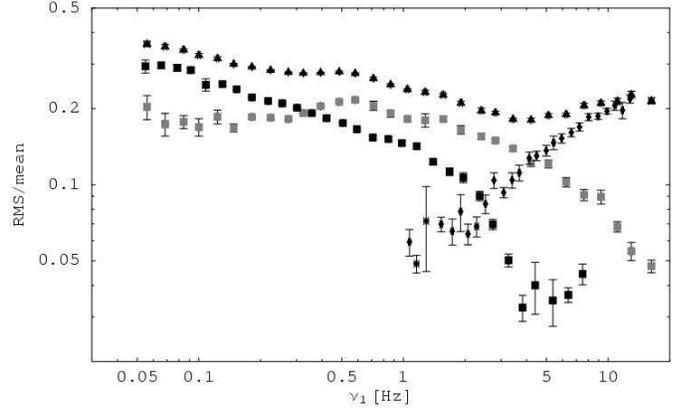}}
\caption{Total fractional RMS (triangles) in the 0.01-50~Hz range and
contribution of each component, plotted against $\nu_1$. Black squares
indicate the contribution from $L_1$ and $L_1^+$, grey squares from
$L_2$ and diamonds from the power-law. Note how the total fractional
RMS stays nearly constant throughout the hard state, dips during the
transitions and increases again in the soft state. The increase below
\mbox{$\nu_1 \sim 0.2$~Hz} corresponds to the observations with highest 
hardness ratio, where a third Lorentzian component is seen to enter
the PDS. These cases make up less than 10\% of the total hard state
observations.}
\label{comprms}
\end{figure}

We now return to the observations, and with the improved results 
from the fits of the soft state, we attempt to separate the power spectral
components and study their evolution through the spectral states. For
this analysis we use both the data presented here, complemented 
with the data of \citetalias{abl05}. The goal should be to ultimately 
understand the
relation between the components of the temporal variability and those
of the radiation spectrum. The transitions are of key importance since
they are associated with many changes in both timing and radiation
properties. To measure the relative strength of each PDS component,
and its evolution, we computed the fractional RMS by integrating over
the $0.01-50$~Hz range. The
result is shown in Fig.~\ref{comprms} together with the total fractional 
RMS, $\widetilde{\sigma}_{\rm tot}$, defined as  
\begin{equation}
\widetilde{\sigma}_{\rm tot} \equiv \frac{\sigma_{\rm tot}}{F} \, ,
\end{equation}
\noindent
where $\sigma_{\rm tot}$ is the total RMS and $F$ the mean flux of
the lightcurve used in calculating the PDS. We furthermore assume 
that
\begin{equation}
\sigma^2_{\rm tot} =\sigma^2_{\rm lor} + \sigma^2_{\rm pl} \, ,
\label{variances}
\end{equation}
\noindent
where $\sigma_{\rm lor}$ and $\sigma_{\rm pl}$
correspond to the RMS of both Lorentzians and the power-law component 
respectively. It is assumed that both Lorentzians are produced 
by the same process, independent of that generating the power-law component.
Since we neglect power below $0.01$~Hz, the RMS values for the
power-law component should be seen as lower limits.

Figure~\ref{comprms} shows that $\widetilde{\sigma}_{\rm tot}$
remains quite constant throughout the hard state ($\nu_1\la0.65$~Hz). 
The exception to this is the range with the highest hardness ratio,
i.e. at the lowest frequencies ($\nu_1 \la 0.2$~Hz). In these
observations, there is evidence of a third Lorentzian component
entering the frequency window, thus increasing 
$\widetilde{\sigma}_{\rm tot}$ \citep[see, e.g., ][]{now00,pot03}. In 
the more common hard state, $\widetilde{\sigma}_{\rm tot}$ reaches 
a plateau (\mbox{$0.25 \la \nu_1 \la 0.65$~Hz}) with a slope of 
$(-0.003 \pm 0.004)~{\rm Hz}^{-1}$ (zero within $1\sigma$, 800 data
points). However, the contribution from each Lorentzian component is
variable. As the power-law component enters the window, the Lorentzian
components weaken. Although the power-law component steadily increases
in power, it does not fully compensate for the weakening and 
$\widetilde{\sigma}_{\rm tot}$ declines during the transition. In the soft 
state the power-law starts to dominate the PDS and 
$\widetilde{\sigma}_{\rm tot}$ once again increases.

\begin{figure}
\resizebox{\hsize}{!}{\includegraphics{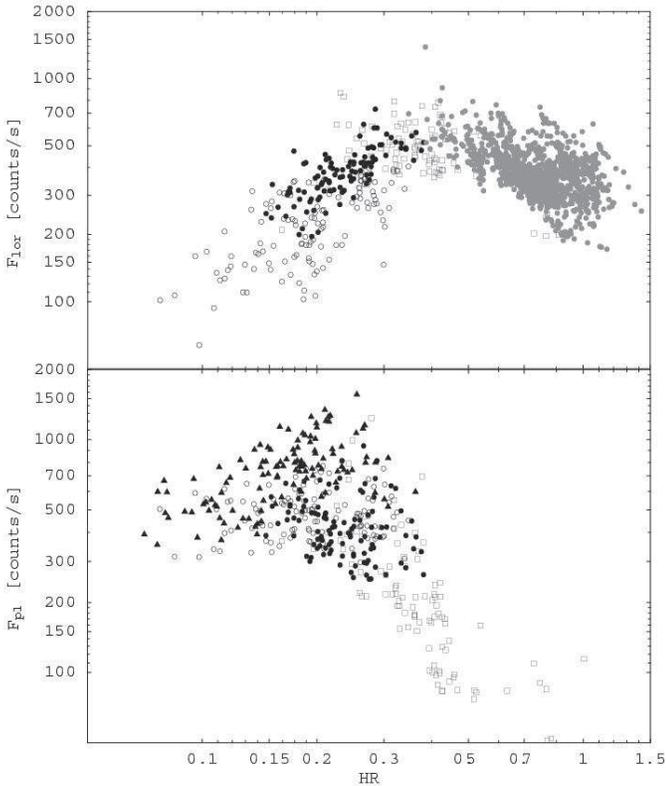}}
\caption{Flux relating to the Lorentzian and power-law components as a 
function of hardness (9-20 keV over 2-4 keV) under the assumption that
the combined fractional RMS of the Lorentzians is constant. The
Lorentzian components (upper panel) follow roughly the same
hardness-flux pattern as the source in general \citepalias[compare,
e.g., Fig.~7 of ][]{abl05}. However, the monotonic decrease of $F_{\rm
pl}$ (lower panel) with increasing hardness can be explained by the
contribution of the power-law decreasing. The true hardness-flux 
behavior associated with the power-law is positive, and that is why the 
spread of points appears to systematically decrease. The symbols 
refer to the
different model components used: two Lorentzians (grey points), two
Lorentzians and power-law (open squares and black points for
transitional and soft state points respectively), Lorentzian plus
power-law (open circles) and only power-law (filled triangles).}
\label{compev}
\end{figure}

\begin{figure}
\resizebox{\hsize}{!}{\includegraphics{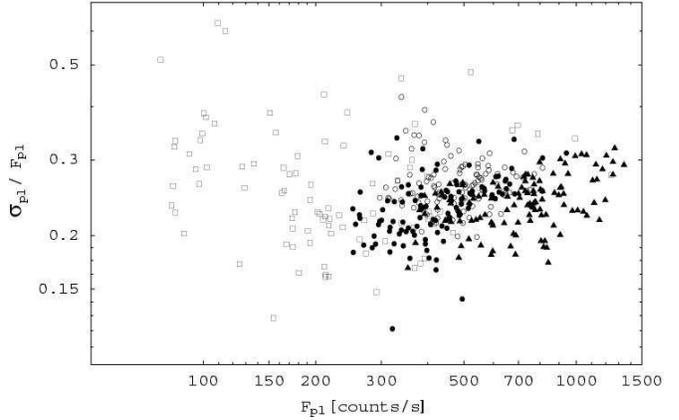}}
\caption{Fractional RMS of the temporal power-law component versus flux
related with the component. There is considerable spread, but the result 
rules out any strong correlation. The symbols are the same as in 
Fig.~\ref{compev}.}
\label{rmsfluxpl}
\end{figure}

Continuing with the assumptions above, we may consider that the components 
modulate two independent emissions so that the flux $F$ 
can be written as
\begin{equation}
F = F_{\rm lor} + F_{\rm pl} \, ,
\label{fluxes}
\end{equation}
\noindent
where $F_{\rm lor}$ and $F_{\rm pl}$ are the fluxes related with 
respectively the Lorentzian and power-law components. However, to derive 
each flux component we need an additional
assumption. As noted above, the fractional RMS of the combined
Lorentzian components remains fairly constant during the hard state
($\simeq 0.28$ with a relative dispersion of 4.4\%), and shows a
decline just when the power-law component appears in our frequency
window in the transitional state. We now assume that
\begin{equation}
 \sigma_{\rm lor} / F_{\rm lor} = const. \,
\label{lor}
\end{equation}
\noindent
for all states, with $\sigma_{\rm lor}$ being the RMS of both
Lorentzians. Thereby associating a smaller fractional RMS to
the power-law component, we can determine the individual flux
contributions in Eq.~\ref{fluxes}. These quantities are shown plotted
versus hardness in Fig.~\ref{compev}, and the RMS of the power-law
($\sigma_{\rm pl}$) is plotted versus flux related with that
component in Fig.~\ref{rmsfluxpl}.

As can be seen by comparing Fig.~\ref{compev} with the total 
flux-hardness relation in Cyg~X-1 
\citetext{e.g., Fig.~7 in \citetalias{abl05}; Fig.~4 in \citealt{zdz02}}, 
it is the behavior of the
Lorentzian components that appears to determine the flux-hardness
correlation. Furthermore, the flux related with the Lorentzians is
reduced in the soft state. The weakening of the Lorentzians in the
soft state is thus due to an actual weakening of the corresponding 
emission component, and not simply due to another emission component
growing stronger. The flux related to the power-law component
decreases monotonically with increasing hardness. Note however that
the spread also increases systematically with flux. As the
hardness-flux relation is known to be positive in the soft state
\citep[e.g.,~][]{wen01,zdz02}, it is likely that the general increase in
flux is merely a result of the power-law component beginning to
dominate the PDS, and that the true hardness-flux relation is
positive, and the cause of the systematic increase in spread. The fractional
RMS of the power-law (Fig.~\ref{rmsfluxpl}) does not show any strong
trend with the associated flux, and the results are consistent with a
constant ratio. As seen in the figure, the dispersion is greatest when
the power-law component is weak, and decreases as the component
becomes stronger. We have calculated the median values in intervals,
and they all coincide (within $2\sigma$) with the median when the
power-law is the only component seen in the PDS (black triangles in
Fig.~\ref{rmsfluxpl}). This value is \mbox{$0.24\pm0.004$}, and the 
sample standard deviation is 0.03 (14\%).

\section{Discussion}
\label{discuss}

In order to establish whether a shift of $L_1$ to $L_1^+$ is truly
a possible scenario, all parameters must be examined. We also discuss 
the results of the PDS decomposition, which provide clues to the relation 
between the components.

\subsection{Shift to first harmonic}

If the apparent change in the frequency correlation is caused by $L_1$
shifting to a higher harmonic, this should be visible in the other
parameter correlations as well. The behavior of $H_1$, $W_1$ and
hardness ratio as a function of fitted Lorentzian frequency does
indeed show some clear changes during the transition.

As noted above, the value
of the width parameter $W_1$ appears to center around 1.2 in the hard
state and another, lower value of $\sim 0.7$ in the soft state (panel
\textbf{b} in Fig.~\ref{parrels}). While this can simply indicate that
$L_1$ becomes more defined in the soft state, it is interesting to note 
that a shift to twice the frequency also predicts an increase in the
coherence of the signal by a factor of two. While $W_1$ is not
directly a measure of the coherence, the decrease is consistent with 
the higher coherence expected if observing $L_1$ in the hard state and 
$L_1^+$ in the soft state. Our parametrization does 
not directly give a value for the more generally quoted quality factor 
$Q=\nu/\Delta \nu$. The relation between $W$ and $Q$ is:
\begin{equation}
W=2\left[ 1- \left( 1+\frac{1}{4Q^2}\right)^{-\frac{1}{2}}\right].
\label{qwrel}
\end{equation}

Another clue may be obtained when looking at the behavior of $H_1$ (panel 
\textbf{a} in Fig.~\ref{parrels}). 
There is considerable spread in the hard state, but as the source
approaches the transition, the value appears to increase slightly,
only to decrease sharply as the transition is reached. In the context
of a shift to the first harmonic, the rise would start when the
strength of $L_1^+$ becomes comparable to that of $L_1$, and the rapid 
decline is seen when $L_1$ begins to weaken. As noted in Sect.~\ref{sims}
above, this has to happen fairly rapidly, consistent with the sharp
drop in $H_1$.

\begin{figure}
\resizebox{\hsize}{!}{\includegraphics{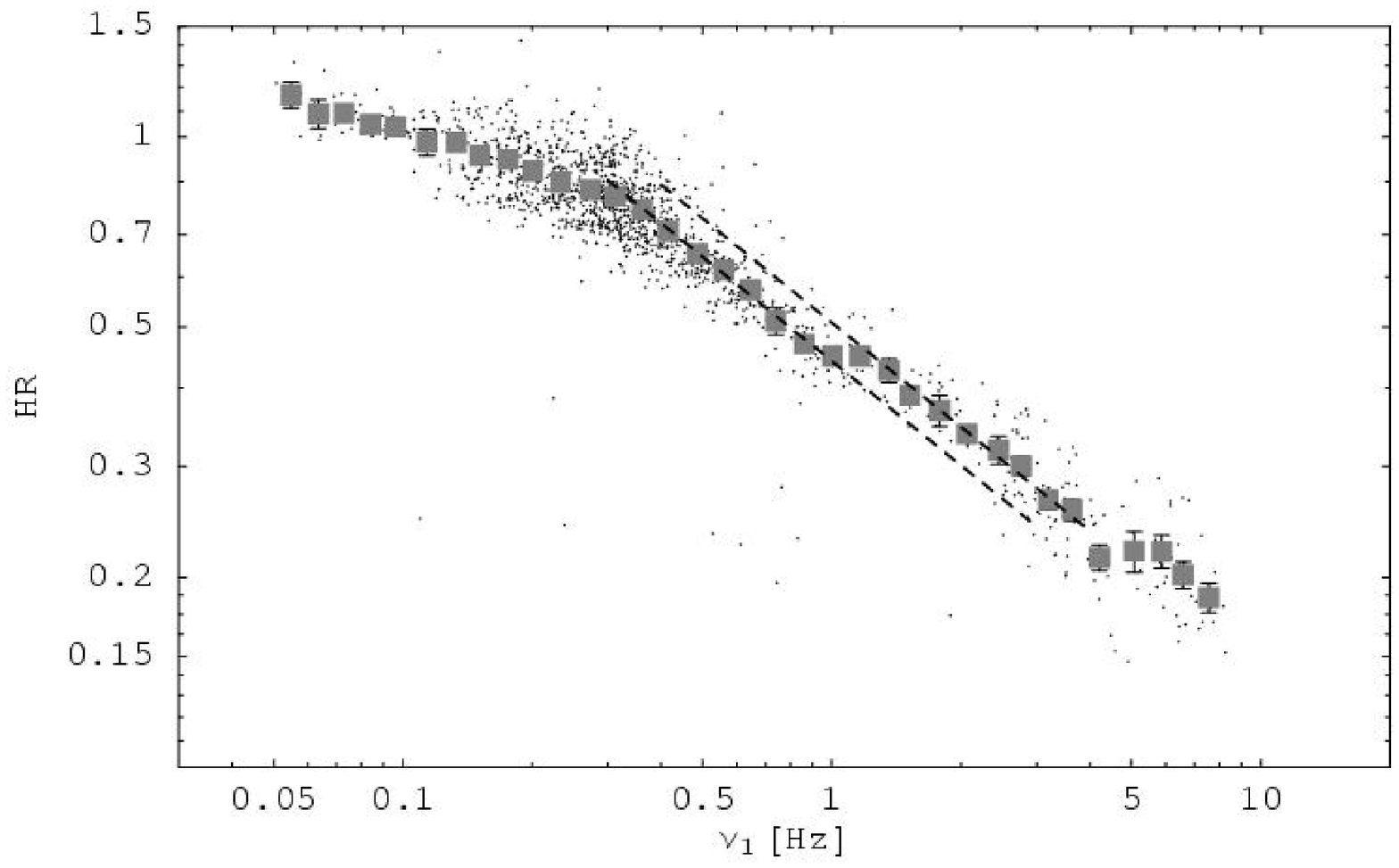}}
\resizebox{\hsize}{!}{\includegraphics{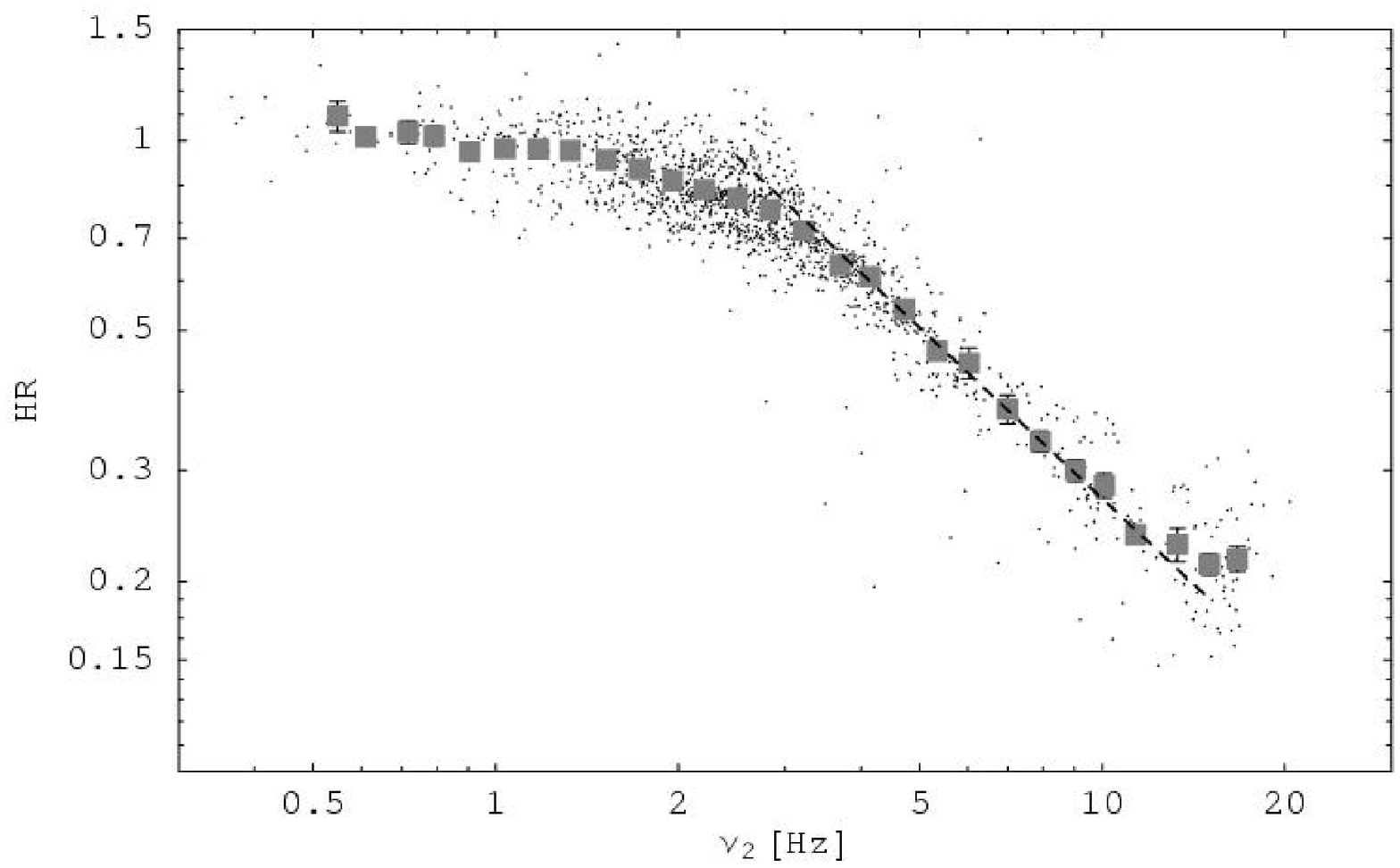}}
\caption{Hardness ratio as a function of $\nu_1$ and $\nu_2$
respectively. While the relation appears fairly smooth for both
components, there is evidence of a shift in the $\nu_1$ relation. The
filled squares show the results when binning the points along frequency.
The dashed lines show the behavior of the $\nu_i$-HR relation. For $\nu_2$
(lower panel), a single power-law describes the relation through all the
states. However, for $\nu_1$ it is necessary to shift the power-law
during the transition, as seen in the upper panel.}
\label{evidence}
\end{figure}

As the hardness is not a parameter of either Lorentzian component, it
gives an independent measure of the evolution of the source. It also
allows a comparison to determine if the shift in the frequency
relation is related to changes in $\nu_1$ or $\nu_2$ (or
both). Plotting the hardness against these two parameters
(Fig.~\ref{evidence}) does reveal some differences.  For $\nu_2$, the
relation is a smooth power-law, with a flattening at the highest
hardness. However, while the $\nu_1$ relation also describes a
power-law, there is what appears to be a shift during the transition.
Though not immediately evident, this shift indicates that the change
in the frequency relation is indeed caused by changes in $\nu_1$, and
is furthermore what would be expected if a shift to the second
harmonic occurs.

Taken together, the above features provide rather convincing evidence
for the idea of a shift from $L_1$ to $L_1^+$ during the
transition. The observed properties are certainly consistent with that
interpretation. Any suggested model needs to explain all the features
consistently, and we find that invoking the shift to $L_1^+$ is a
simple but successful scenario. 

We also note that in their long-term study of the PDS of Cyg~X-1, 
\citet{pot03} report a shift in power between their Lorentzian
components. In the hard state, the PDS are dominated by $L_1$ and
$L_2$, while during the transitions and `failed state transitions'
$L_1$ decreases in power and $L_3$ strengthens. As their identification
of components is not the same as that used here, a direct comparison is 
not possible. However, the mechanism suggested above is consistent
with their results. In addition, results by \citet{pot05} from  
intermediate state PSD indicate that $L_1$ may decrease earlier at higher
energies.  

An appearance of a QPO with harmonic content has been reported by
\citet{bel05} in GX~339-4 in connection
with the source shifting from the hard state to the intermediate state.
If the change in the frequency relation observed in Cyg~X-1 
during state transitions is indeed caused by $L_1^+$ growing in power, 
the question arises if the appearance of a harmonic is a more general 
feature of the intermediate state of black hole binaries. 
The appearance of (or in the case of Cyg~X-1, a gradual shift to) a harmonic 
could then be used as an indicator of spectral state. Certainly the 
change in the frequency relation may used as a gauge of the state 
transition in Cyg~X-1, and may as such be added 
to the growing list of possible ways to determine the state of the source.
It is likely that all these indicators are responding to the same changes
in the source (e.g., accretion geometry) rather than being directly related.
Nonetheless, when combined they do provide constraints on proposed models 
for the state transitions.

In the context of the relativistic precession model discussed in
\citetalias{abl05} \citep{SV98,SV99,SVM99} $\nu_1=2 \nu_{\rm nod}$ and
\textbf{$\nu_2=\nu_{\rm per}$}, where $\nu_{\rm nod}$ is the frequency of
nodal precession and $\nu_{\rm per}$ the frequency of periastron
precession. The shift would then mean that $\nu_1^+=4 \nu_{\rm
nod}$. Assuming the frequencies are produced close to the inner edge
of an accretion disk \citep[or picked out, e.g., from a narrow
transition region, see ][]{psa00}, extending the relation to higher
frequencies allows tracking of the disk to smaller radii. Assuming a
prograde scenario with dimensionless specific angular momentum $a_* =
0.49$,
\begin{equation}
\frac{R_{\rm in}}{R_g}\simeq 16.5 \left(\frac{M}{8\;{\rm M_\odot}}\right)^{-\frac{2}{5}} \left(\frac{\nu_{\rm per}}{10\;{\rm Hz}}\right)^{-\frac{2}{5}},
\end{equation}
\noindent
where $R_{\rm in}$ is the inner disk radius, $R_g$ the gravitational
radius, and $M$ the black hole mass. The dependence on $\left|a_*\right|$
is very weak in the studied range.

\subsection{Decomposed power spectrum}

Although the RMS of the two Lorentzian components show different behaviors
and evolution during the hard state, the sum remains constant. While this
is no direct evidence that the components are tied, it is improbable
that two unrelated components would combine to a constant total. We therefore
view the behaviors of the Lorentzians as indicative of them arising from 
the same physical process. One possibility (especially in the context of 
the relativistic
precession model discussed above) is a scenario where the two components 
share the same, constant energy. As the amplitude of one grows, the energy 
available to the other will drop and vice versa. 

An extension of this reasoning indicates that the variability giving rise
to the power-law component is instead independent of the process behind
the Lorentzians. When the power-law enters the frequency window, the total
RMS drops, reaching a minimum when the PDS components are all roughly of
the same strength. As shown by \citet{zdz05}, this is precisely what 
is expected if the measured flux arises in different and unrelated 
processes. We therefore conclude that there are at least two unrelated
radiation components producing the PDS of Cyg~X-1. One dominates the hard
state and provides the variability seen in the Lorentzian components, the 
other dominates in the soft state and produces the power-law component.
The drop in total RMS during transitions can also be seen in Fig.~3 
of \citet{pot03}.

This scenario fits with previous analyses of the radiation spectrum,
which typically include both a soft black-body component and a harder
component usually ascribed to Comptonization in a hot corona
\citep[e.g., ][]{gie97,gie99}. Recent studies of the energy-dependent
variability of Comptonization
\citep{gie05} show promising results in tying together temporal and 
spectral components. Work along these lines has also been presented
by \citet{miy94} for the source \object{GS~1124-683}, and more recently 
by \citet{vig03} and \citet{zyc05}. 

An interesting possibility for the two components has been suggested
by \citet{ibr05}. They find that the radiation spectrum of the hard state
of Cyg~X-1 can be well fit using both a thermal and non-thermal
distribution of electrons, associated with an inner hot corona and
magnetic flares above the accretion disk respectively. The transitions 
into the soft state can then be explained by the inner edge of the disk 
moving to smaller radii. As this happens, the corona is reduced (thus
weakening the thermal Comptonization component), and emission from the
non-thermal distribution connected to the flares becomes more
prominent in the radiation spectrum. It is tempting to associate the
Lorentzian components with a transition region between the cool disk
and the inner quasi-spherical corona, and similarly tying the
power-law component of the PDS to emission from a non-thermal distribution 
of electrons, associated with the flares above the disk. Not only does 
the relative strength of the power-law
increase, but also the cut-off moves to higher frequencies as the
source enters the transitional and soft states, when the disk is
expected to extend closer to the compact object. Comparing the
results of this study with those gained from fitting the radiation 
spectrum of Cyg~X-1 in all states \citep[e.g.,][]{ibr05,wil05} indicates that 
there is no trivial, direct connection between the components of the 
radiation spectrum and those of the PDS. In order to build a complete 
physical picture, the behavior of the radio jet of the source also 
needs to be included. Studies have found correlations between the radio 
and X-ray flux in Cyg~X-1 \citep[e.g.,][]{bro99,poo99}, and the radio flux
has been connected to the state transitions \citep{wil05}. Furthermore, 
\citet{mig05} have found correlations between radio luminosity 
and temporal features of the PDS in a number of X-ray binaries.

If the power-law PDS component is taken to have constant fractional
RMS, the value cannot be the same as that for the Lorentzian
components.  While the distribution in Fig.~\ref{rmsfluxpl} does not rule
out a trend in the fractional RMS, drastic changes are not consistent
with the data. We have performed Spearman rank correlation
analysis and find only very weak correlations between the parameters
of the power-law and those of the Lorentzians. If the power-law
component is connected with simple shot noise, the flat behavior seen
in Fig.~\ref{rmsfluxpl} indicates that the increase in flux is due to
the amplitude of the shots increasing, and not to a greater number of
shots, as that would reduce the observed fractional RMS. The results
are consistent with the RMS being proportional to the flux, as shown
by \citet{um01}.

In a study of the relation between RMS and flux on both longer and
shorter timescales, \citet{gle04} find that the fractional RMS in the $1-32$
Hz frequency band to be constant throughout the hard state changes of PDS, 
similar to the results presented above. They investigate the RMS-flux 
correlation in terms of a
linear relation and show that such a relation is present in both the hard
and soft states of Cyg~X-1. However, the slope and intersect of the linear
relation have different values in different states. While the paper mainly 
concentrates on interpretations 
in terms of a component with either constant flux or constant RMS, the authors
also mention the possibility of a component with constant fractional RMS. 
We find this interpretation to fit well with the results presented in this
paper, where both the sum of the Lorentzian components as well as the 
power-law component display such behavior. As noted above, the two cannot
have the same value for a constant fractional RMS, and it is therefore 
natural that the parameters the RMS-flux relation be different in the hard 
and soft state, as the power-law is not visible in the former but dominates
the latter. During the intermediate state, both components are present, 
resulting in parameter values in-between the hard and soft state values,
as reported by \citet{gle04}.

\section{Conclusions}
\label{conc}
We have conducted an in-depth investigation of the soft and transitional
state PDS of
Cyg~X-1 both reanalyzing previous data and adding data from new soft 
state observations. Fitting the PDS with a model of two Lorentzian
components and cut-off power-law, we are able to extend the frequency
relation \citetext{\citealt{wvdk99}; \citetalias{abl05}} into the soft 
state. We show that the index of the relation is consistent with that 
found in the hard state. This further supports the identification of
the frequencies as the relativistic precessional frequencies.

Using an approach of Monte Carlo simulations we show that a shift in
$L_1$ from $\nu_1$ to its first harmonic $\nu_1^+=2\nu_1$ is consistent 
with the behavior of all parameters. We conclude that this scenario provides 
a simple explanation of the observational data.

Combining these results with the hard and transitional state results
from \citetalias{abl05}, we decomposed the PDS and studied the evolution
of the individual components. The results show that although each
Lorentzian is variable, the total fractional RMS is constant throughout 
the hard state of the source. In contrast, when the power-law component 
enters the frequency window, the total RMS drops. Together with the lack 
of correlation between power-law and Lorentzian parameters, this suggests 
that these components arise in different regions and/or through different
processes. Using the observed behavior, we attempted to derive the
relation between hardness and flux for the PDS components, as well
as the relation between flux and RMS for the power-law. We conclude that
the weakening of the Lorentzian components as the source enters the 
soft state is due to an actual decrease in the associated flux component. 

The results show that at least two variability components must be
present in the 2-9 keV energy range studied here. In order to better
understand both Cyg~X-1 and similar objects, studies of both temporal
and radiation spectra need to be taken into account. Only by combining
the insights from these fields can we hope to understand these
sources.

\acknowledgements
We are grateful to the anonymous referee for constructive comments 
which helped improve the paper, and we also wish to thank Linnea 
Hjalmarsdotter and Juri Poutanen for stimulating discussions. This 
research has made use of data obtained through the
High Energy Astrophysics Science Archive Research Center (HEASARC)
Online Service, provided by NASA/Goddard Space Flight Center.

\end{document}